%% file: main.tex
\documentclass[nonacm,sigplan]{acmart}

\usepackage[normalem]{ulem}
\hypersetup{colorlinks={true},linkcolor={blue},citecolor=blue}
\usepackage{cleveref}
\crefname{section}{}{\S\S}
\crefdefaultlabelformat{\S#2#1#3}
 
\usepackage{xspace}
\usepackage{comment}
\usepackage[normalem]{ulem}
\usepackage{enumitem}
\setlist[enumerate]{itemsep=0mm}
\usepackage{multirow}
\usepackage{float}
\usepackage{makecell}

\setlist{nolistsep}
\usepackage{url}

\usepackage{listings}
\usepackage{color}
\usepackage[font=footnotesize]{caption}
\usepackage[font=footnotesize]{subcaption}
\usepackage{tabularx}

\usepackage{textcomp}
\usepackage{booktabs}

\newcommand{\design}{\textsc{Rinas}\xspace}
\begin{document}

% \title{
% \textsc{Undertale}\xspace: Accelerating Model Training via Unordered Batch Generation for Large Shuffled Datasets
% }

\title{
 \textsc{Rinas}\xspace: Training with Dataset Shuffling Can Be General and Fast
}

\author{Tianle Zhong}
\email{tianle.zhong@email.virginia.edu}
\affiliation{
\institution{University of Virginia}\country{}
% \city{Charlottesville}
% \state{Virginia}
}

\author{Jiechen Zhao}
\email{jc.zhao@mail.utoronto.ca}
\affiliation{\institution{University of Toronto}\country{}}

\author{Xindi Guo}
\email{ptk2ks@virginia.edu}
\affiliation{\institution{University of Virginia}\country{}}

\author{Qiang Su$^{\ast}$}
\email{qiang.su@my.cityu.edu.hk}
\affiliation{\institution{City University of Hong Kong}\country{}}

\author{Geoffrey Fox$^{\ast}$}
\email{vxj6mb@virginia.edu}
\affiliation{\institution{University of Virginia}\country{}}

\thanks{$^{\ast}$Qiang Su and Geoffrey Fox are corresponding authors. The Univesity of Virginia team thanks NSF Grant 2210266 and  DOE Grant DE-SC0023452 for partial support. We acknowledge the excellent work of the Rivanna HPC Cluster team. }

% \begin{acks}
%     The Univesity of Virginia team thanks NSF Grant 2210266 and  DOE Grant DE-SC0023452 for partial support. We acknowledge the excellent work of the Rivanna HPC Cluster team.
% \end{acks}

% \affiliation{%

\begin{abstract}
Deep learning datasets are expanding at an unprecedented pace, creating new challenges for data processing in model training pipelines. A crucial aspect of these pipelines is dataset shuffling, which significantly improves unbiased learning and convergence accuracy by adhering to the principles of random sampling. However, loading shuffled data for large datasets incurs significant overhead in the deep learning pipeline and severely impacts the end-to-end training throughput. To mitigate this, current deep learning systems often resort to partial dataset shuffling, sacrificing global randomness to maintain acceptable training throughput on large datasets, still leaving global shuffling efficiency issues not fully explored.
%Existing efforts seek to balance and schedule such partial shuffling for certain deep learning workloads to achieve higher training throughput

In this work, we present \design{}, a data loading framework that systematically addresses the performance bottleneck of loading global shuffled datasets. Our key contribution is to offer an intra-batch unordered data fetching approach, which unleashes unexplored parallelism of data loading. We implement \design{} under the PyTorch framework for common dataset libraries HuggingFace and TorchVision. Our experimental results show that \design{} improves the throughput of general language model training and vision model training by up to 59\% and 89\%, respectively.
\end{abstract}

\maketitle % should come after the abstract
\pagestyle{plain} % should come right after \maketitle

\input{content/introduction}

\input{content/background}

\input{content/motivation}

\input{content/design}

\input{content/implementation}

\input{content/evaluation}

\input{content/discussion}

\bibliographystyle{plain}
\bibliography{references}

\end{document}

%% file: content/introduction.tex
\section{Introduction}

Shuffling large datasets remains a critical issue in data processing systems, impacting a wide spectrum of applications \cite{dean2008mapreduce, 10.1145/2934664, nicolae2016towards, shen2020magnet, hadoop}. In the realm of deep learning, dataset shuffling is not merely a procedural task but a fundamental aspect of the data loading pipeline. It is instrumental in avoiding overfitting better \cite{ying2019overview, li2019research} and benefiting convergence accuracy based on the theoretical foundation \cite{meng2019convergence}.

Ideally, the training accuracy improvement derived from data shuffling should introduce minimal shuffling overheads on overall training throughput. 
Unfortunately, as the sizes of deep learning datasets expand rapidly, the shuffling overhead is significantly growing. 
This is because the datasets nowadays are towards tens of TBs, largely exceeding the system DRAM capacity and thus leading to the use of slower disk I/O becoming an inevitable bottleneck~\cite{lobster, sun2022solar, gu2022fluid}.
Worse, it is challenging for current systems to effectively manage the shuffling overhead on training throughput without sacrificing accuracy~\cite{nguyen2022globally, exoshuffle}.

The impact of large dataset shuffling on training throughput is substantial and far from optimal. For example, we observe that data loading I/O for shuffling can consume up to 85\% of the total training time for models such as ResNet-152 on the ImageNet dataset ($\sim$140 GB). This phenomenon corroborates with the prior art~\cite{10.1145/3458817.3476181}. For language model training, the slowdown caused by shuffled loading can also lead to 30\% to 50\% of training throughput degradation. This overhead persists as a dominant bottleneck in training efficiency, even when deep learning pipelines are integrated with advanced data processing systems and databases~\cite{deeplake}.

In the practice of typical deep learning systems, shuffling for larger-than-memory datasets is usually abstracted by a shuffled loading operation, fetching a batch of data samples randomly from the dataset in a on-demand fashion, which avoids pre-loading the entire dataset into DRAM for shuffling.
To fetch the data from the disk to the system DRAM, each data sample needs to be indexed inside the dataset which enforces the data loading to perform random disk I/O operations \cite{Index, hf_map_vs_iter}.

Amidst this backdrop, contemporary research has made strides in two directions: (1) refining the shuffled dataset loading pipeline to conceal its impact on end-to-end training throughput; and (2) identifying an equilibrium between shuffling thoroughness and converged accuracy. However, these advancements come with concessions: (1) hiding data loading overhead is ineffective when it becomes the predominant factor in training performance; (2) the complexity of the data loading pipeline demands extensive modifications to existing software infrastructure; (3) the equilibrium point is not universally applicable across the spectrum of datasets and learning models; (4) the scope of trade-offs is constrained by system capabilities, leaving practitioners to contend with a compromise that sacrifices either accuracy or speed, since a not thorough shuffle harnesses the convergence accuracy \cite{xu2022stochastic}.

The fundamental question is, can we devise a universal framework that accelerates the loading of large, shuffled datasets without compromising accuracy and training speed across diverse datasets and learning models?

To address this problem, we introduce \design{}, a comprehensive data loading framework designed for efficient model training on large shuffled datasets. 
We pinpoint the frequent random disk I/O on individual data samples as the principal bottleneck in loading shuffled datasets. 
Under this observation, we propose a novel data preparation method for model training processes: intra-batch unordered data fetching.

\design{} is predicated on a key insight: within a training iteration, the sequence of computing the average loss from a batch of randomly sampled data does not influence the learning outcome. This insight suggests that the data retrieval order for intra-batch samples does not affect the learning process. 
Fundamentally, it allows us to shift from a strict global sample order to a more flexible intra-batch unordered manner. 

This insight brings in several benefits. First, relaxing such orders enables parallel data retrieval without stringent ordering constraints.
Second, \design{} fundamentally negates the need to balance between accuracy and speed. 
Third, \design{} is versatile enough to cater to various datasets and learning models because of the guarantee of equivalent learning outcomes.

We architect \design{} with a data-agnostic 
control plane and a data plane for on-demand and parallelized data fetching. 
The control plane offers an execution model for loading samples and generating batches with them in an unordered fashion. 
The data plane fully exploits the advantages of unordered parallel data retrieval across diverse dataset structures.

Our implementation showcases \design{}'s applicability across major learning tasks from computer vision to language models. 
We demonstrate our prototype within the PyTorch DataLoader \cite{Index}, as well as TorchVision \cite{torchvision2016} and HuggingFace Datasets libraries \cite{lhoest-etal-2021-datasets}.
Our experiments evaluate standard training workloads on large datasets, spanning from computer vision model training to language model pretraining, within typical deep learning training clusters. 
Our assessments cover the typical training setup, revealing \design{}'s minimal loading-related overheads at different scales, delivering up to 59\% and 89\% speed increases in training for computer vision and language models, respectively.

The ensuing sections will delve into the background (\S \ref{sec: background}) and motivation (\S \ref{sec: moti}), review related work and its limitations, and then articulate the design (\S \ref{sec: design}) and implementation (\S \ref{sec:impl}) of \design{}. Finally, we will present our evaluation findings (\S \ref{sec: eval}) and engage in a discussion (\S \ref{sec: discussion}).

%% file: content/background.tex
\section{Background}
\label{sec: background}

\begin{table}[t]\small
\resizebox{\columnwidth}{!}{
\begin{tabular}{l|l|l}
\hline
\textbf{Dataset} & \textbf{Size} & \textbf{Description} \\ \hline
ImageNet \cite{deng2009imagenet} & 140 GB & For image classification \\
ImageNet-21k \cite{deng2009imagenet} & 1.8 TB & Extended version of ImageNet \\ \hline
RedPajama \cite{together2023redpajama}  & 5 TB  & \multirow{2}{*}{Language modeling} \\
C4 \cite{2019t5}                        & 7 TB  &  \\ \hline
CosmoFlow \cite{mathuriya2018cosmoflow} & 10 TB & Cosmological simulations \\ \hline
\end{tabular}
}
\caption{Large datasets for model training in computer vision and language modeling.}
\label{table: datasets}
\end{table}

\subsection{Storing Emerging Very Large Datasets on Disks}
\label{subsec: large_ds}

ML model training has witnessed increasingly huge datasets that are beyond the system's memory capacity (DRAM), and it becomes impractical to preload the entire dataset before training. Therefore, data preparation emerges as a critical path for model training: loading large datasets into memory is often non-trivial. Table~\ref{table: datasets} presents typical datasets for computer vision models and large language models. The computer vision dataset collects a huge amount of image files, and accessing these images typically requires indexing into a structured file system; Large language models \cite{devlin2019bert, thoppilan2022lamda, touvron2023llama} are trained on vast text corpora, with huge data volumes and complicated structures \cite{together2023redpajama, 2019t5, OpenOrca}. 
These datasets are often segmented into multiple files, each packed with text entries. Retrieving specific samples requires effective parsing or database systems with sophisticated indexing for efficient search and access \cite{lhoest-etal-2021-datasets, torchvision2016}. Specifically, image datasets are usually composed of individual image files on disk as data samples, while text datasets are usually composed of a series of large files on disk, each containing rows of text as data samples.

\subsection{Dataset Shuffling in Model Training Pipeline}

\begin{figure}[t]
    \centering
    \includegraphics[width=\linewidth]{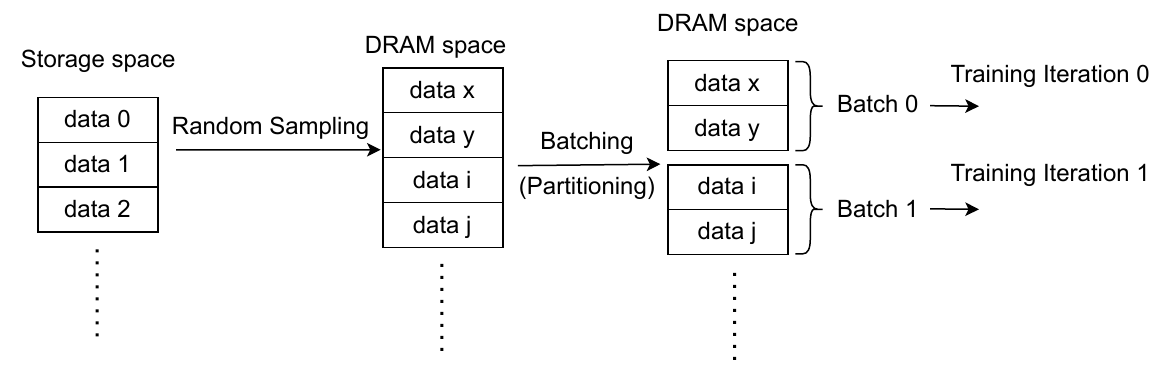}
    \caption{The logical data path of shuffling in end-to-end training.}
    \label{fig: data_path}
\end{figure}

Figure \ref{fig: data_path} presents the typical training pipeline upon the training datasets, where the random sampling~\cite{olken1995random} on the datasets serves as a critical operation. 
The dataset on the storage is randomly sampled into DRAM space, then the sampled data are partitioned into batches to be fed into model training iterations. 

In practice, the operations to achieve random sampling upon the dataset are generally described as \textit{dataset shuffling}~\cite{liu2017shuffle_spark}.
By shuffling the dataset at the beginning of each training epoch, the data sample order to be exposed to the model training process is randomized. Typically, there are two ways of dataset shuffling: shuffling the dataset in the storage space (i.e., in-place shuffling), or loading the dataset into DRAM space with a random order (i.e., shuffled loading).

In-place shuffling is a common operation in large-scale batch processing systems \cite{dean2008mapreduce, hadoop, 10.1145/2934664}. However, unlike smaller in-memory collections, shuffling large datasets that reside on persistent storage devices imposes a considerable overhead \cite{iShuffle, liu2017shuffle_spark}, and involves more than mere in-place reordering. It requires careful consideration of the I/O throughput, storage latency, and computational load on the system  \cite{welton2011improving, zhang2006storage, deeplake, gupta2015amazon}.

Unfortunately, systems highly optimized for in-place shuffling like Hadoop \cite{hadoop} and Spark \cite{spark} prevent reading shuffled results until the shuffle is fully completed, making model training systems hard to pipeline the shuffling process with model training. On the other hand, shuffled loading avoids expensive and complicated in-place reordering of the datasets in the storage space, able to be pipelined with the model training process. As a result, shuffled loading is our paper's focus and many other papers' focus is how to pipeline the in-place shuffling with model training \cite{ray, exoshuffle}. Typically, there are two ways of shuffled loading: Buffered shuffling and indices mapping. They are both popular choices and already supported by many frameworks like PyTorch \cite{Index, ofeidis2022overview}.

\begin{figure}[t]
    \centering
    \includegraphics[width=\linewidth]{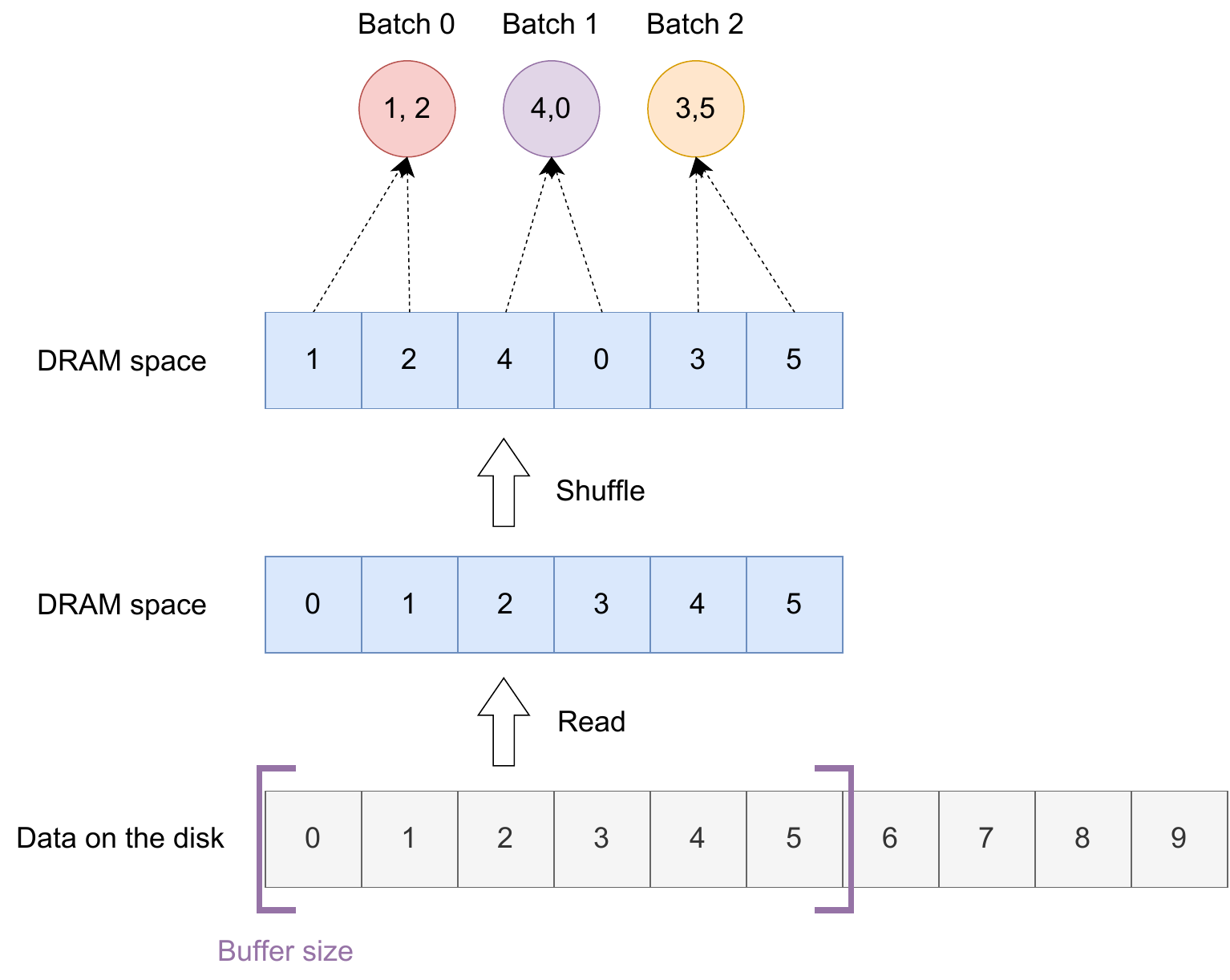}
    \caption{An example of buffered shuffle workflow.}
    \label{fig:buffered_shuffle}
\end{figure}

\noindent{\bf Buffered shuffling.}  Figure \ref{fig:buffered_shuffle} presents an example workflow of buffered shuffling, which involves two steps to improve the efficiency of shuffled loading. First, it leverages \textit{partial shuffling}, sequentially loading a \textit{subset} of the dataset from the disk into a memory buffer allocated in the system DRAM. 
Second, it performs the shuffle operation in this constrained space, followed by the formation of batches from this shuffled subset.

This approach strikes a balance between the need for random access and the performance limitations imposed by disk-based storage, aiming to provide a \textit{partially} shuffled dataset without the prohibitive overhead of random disk I/O \cite{DeepIO_buffer_shuffle}.
However, buffered shuffling cannot achieve true random sampling due to the limited shuffling space compared to global shuffling, which may inadvertently compromise convergence accuracy  \cite{nguyen2022globally, xu2022stochastic}.

\begin{figure}[t]
    \centering
    \includegraphics[width=\linewidth]{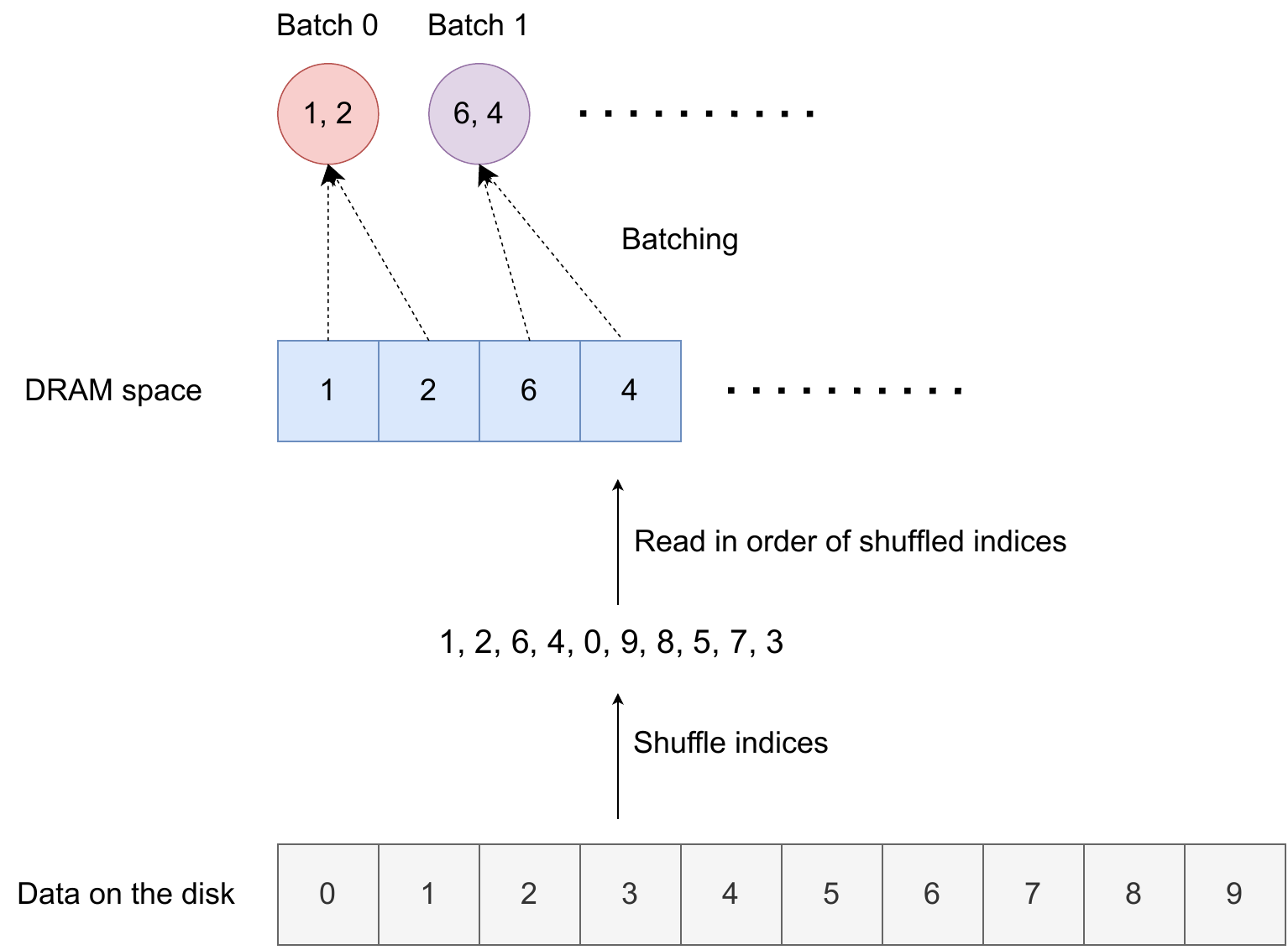}
    \caption{An example of indices mapping workflow.}
    \label{fig:indices_mapping}
\end{figure}

\noindent{\bf Indices mapping.} To avoid the loss of convergence accuracy due to the compromised shuffle quality by partial shuffling, the dataset should be \textit{globally} shuffled \cite{meng2019convergence}. Figure \ref{fig:indices_mapping} depicts an example of indices mapping workflow. In this strategy, the dataset indices are shuffled rather than data, and the data is read into DRAM following the order of shuffled indices.

This approach successfully performs a shuffled loading strategy that fully respects the random sampling principles but creates a sequence of random IO that the storage system is typically less efficient at.

Since indices mapping guarantees true random sampling to benefit model training convergence accuracy at most, our work focuses on addressing the performance issue of indices mapping.

%% file: content/motivation.tex
\section{Motivation}
\label{sec: moti}

In this section, we begin with analyzing the current problems that indices mapping incurs. 
Next, we discuss existing solutions and their limitations.

\subsection{Inefficiency with Indices Mapping for Very Large Datasets}
\label{subsec:inefficiency_problem}

As the previous section discussed, although indices mapping reserves the global randomness of the dataset, this approach is typically much slower than buffered shuffling due to the necessity of non-contiguous data retrieval from storage. Next, we explain the relationship between such a slowdown and the dataset size.

\begin{figure}
\centering
\includegraphics[width=\linewidth]{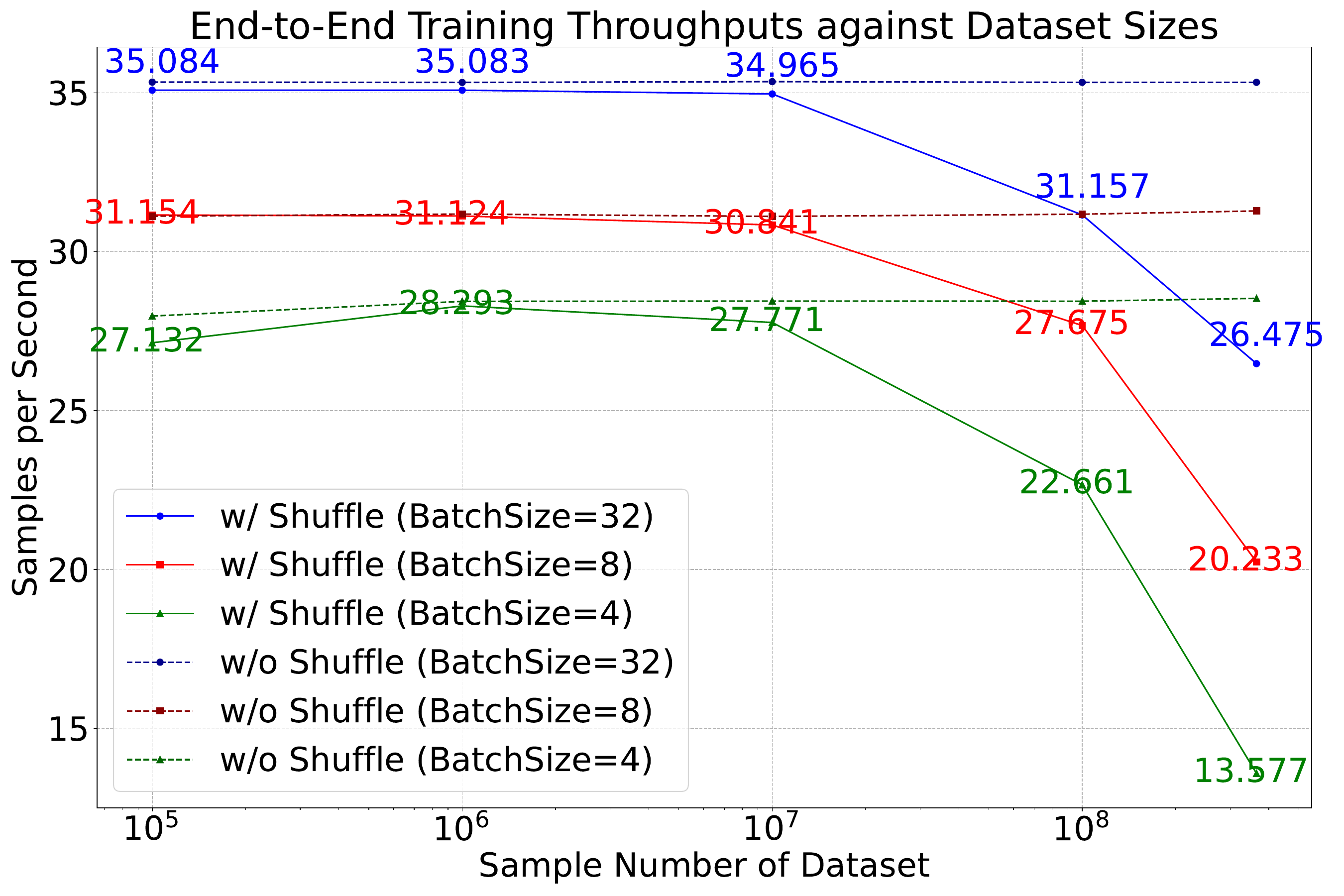}
\caption{Training throughput for the RoBERTa-base model for varying batch sizes within the C4 subsets of different sizes. A discernible trend emerges, showing a decrement in end-to-end training throughput as the number of samples in the dataset increases.
}
\label{fig:train_throughput_moti}
\end{figure}

\noindent{\bf Training slowdown.} 
Figure~\ref{fig:train_throughput_moti} presents the end-to-end training throughput of training RoBERTa-base model~\cite{liu2019roberta} at different batch sizes when dataset size increases on a single NVIDIA A100 GPU. 
While the cleaned English branch of C4 dataset tokenized by RoBERTa has $\sim 3.6\times10^8$ rows,  we synthesize four different sizes of its subsets by choosing its first $10^5$, $10^6$, $10^7$, and $10^8$ rows. We benchmark the training throughput with all five sizes of datasets to show the throughput changes when dataset size increases.
Observe that there is a marked reduction in training efficiency when the dataset size increases, leading to a 30\% to 50\% decrease in speed. 

This reveals that the primary cause of the training deceleration is the overhead associated with shuffling by indices mapping against the large datasets, as opposed to a scenario without shuffling. Notably, the negative impact of this overhead is observed across both large and small batch sizes, underscoring the pervasive influence of shuffling-related delays.

This indicates that in extensive large-scale scenarios, I/O overhead by indices mapping can become the major factor of the total training time, leading to substantial GPU idle time and data starvation. This phenomenon has been corroborated by other research on other training scenarios and systems as well~\cite{10.1145/3458817.3476181, sun2022solar}. 

This also indicates why shuffled data loading overhead for the large dataset cannot be easily hidden by overlapping with computation: under this large extent of degradation, the data loading overhead has dominated the overall training time.

\begin{figure}
\centering
\includegraphics[width=\linewidth]{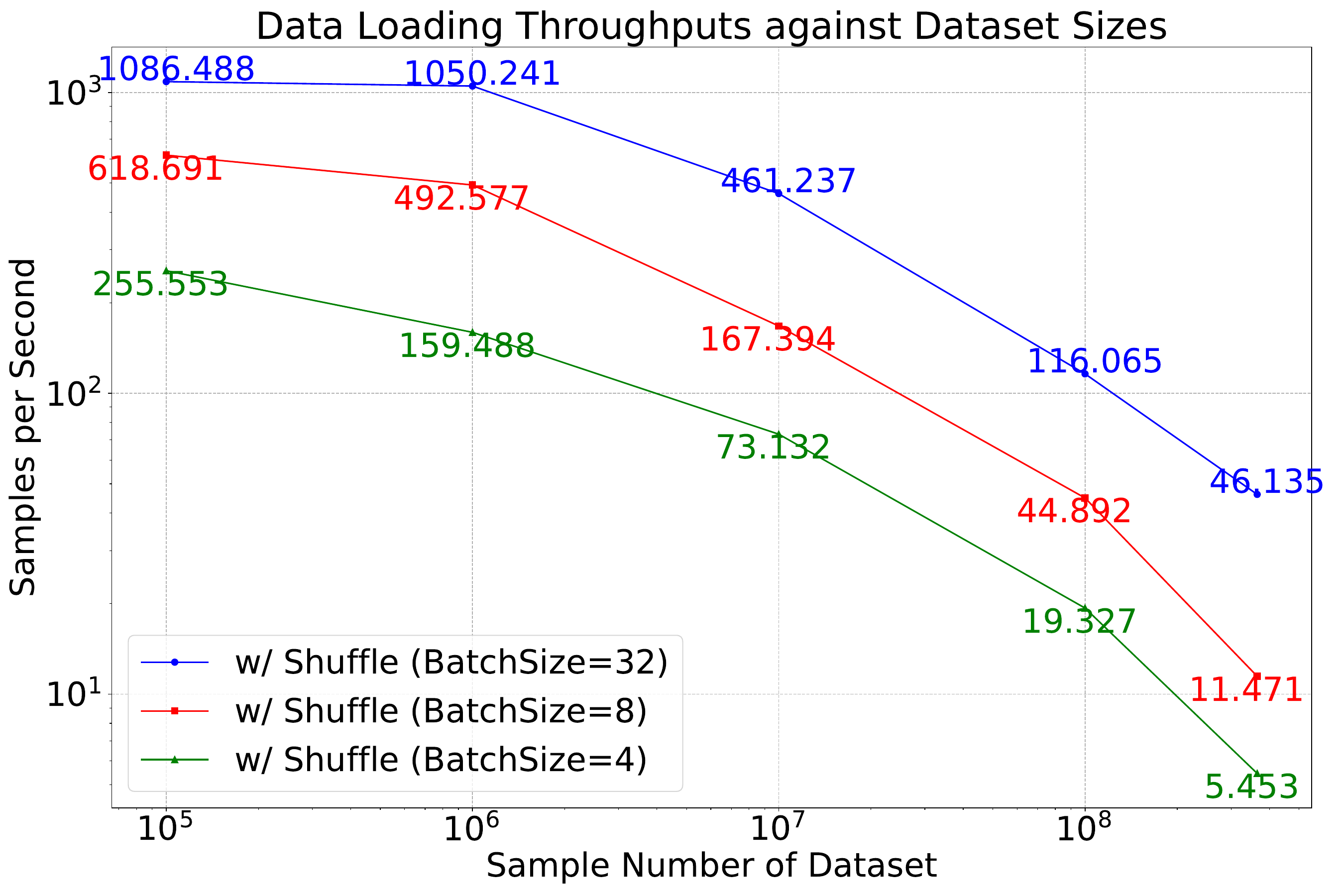}
\caption{Data loading throughput comparison for varying batch sizes within the C4 subsets of different sizes.
}
\label{fig:load_throughput_moti}
\end{figure}

\noindent{\bf Throughput degradation.}  The reason for the training slowdown is shuffled data loading throughput degradation with large datasets. Figure~\ref{fig:load_throughput_moti} presents the pure data loading throughput under the same settings as Figure~\ref{fig:train_throughput_moti_dist} by excluding the model training process and only performing data loading operations.
By further looking into the measured data loading throughput, a parallel decrease in data loading throughput, akin to the trend seen in end-to-end training performance, underscores the impact of data loading efficiency on overall training time.

When employing a batch size of 32, data loading throughput for small datasets using indices mapping can capitalize on system I/O capabilities, achieving a data loading throughput of $\sim$1000 samples per second. However, this throughput significantly diminishes when applied to larger datasets with indices mapping, with performance dropping to $\sim$50 samples per second, primarily due to the random dataset indexing overhead which is positive correlated with the dataset size.

This relationship indicates that throughput degradation during data loading is a primary contributor to the observed training slowdown.
This observation underscores the prominence of data loading as a substantial bottleneck in the training process, particularly when dealing with extensive datasets. This phenomenon necessitates a reevaluation and optimization of data loading practices to alleviate the undue time expenditure and enhance overall training performance.

\subsection{Existing Solutions}\label{motivation:existing-solutions}

Prior work proposes solutions to mitigating the performance issue with large shuffled datasets.
There are three categories of solutions: balanced trade-off between shuffle quality and speed, scheduled data prefetching, and taking advantage of the data-parallel training.

\noindent{\bf Shuffle balancing.} Considering the inefficiency of indices mapping, various research endeavors have experimented with adjusting the degree of shuffling to strike a balance between converged accuracy and training efficiency \cite{nguyen2022globally, sun2022solar}. Nevertheless, the process of identifying this equilibrium can be both time-consuming and highly dependent on the specifics of the model and dataset, limiting its applicability to a wider range of models and datasets. Furthermore, the trade-off space is usually limited by hardware resources which can leave practitioners to contend with a compromise that sacrifices either accuracy or speed \cite{exoshuffle}. For example, for training the ResNet-50 model on the ImageNet dataset, the limited shuffle can result in $\sim$20\% of accuracy drop compared to global shuffled learning \cite{nguyen2022globally}.

\noindent{\bf Data prefetch scheduling.} Given that the sequence of shuffled indices is predetermined, concurrently with the model’s computation on the current batch, the read operation for the next batch can be initiated, allowing data preloading to take place in an overlapped fashion, thereby enhancing performance. There have been efforts such as NoPFS \cite{10.1145/3458817.3476181} and ExoShuffle \cite{exoshuffle} which integrate data loading pipeline with data prefetching scheduling and attempt to overlap data loading with model training computations. Those solutions are very effective when the shuffled loading latency is at a relatively low level such as when employing buffered shuffling or dealing with smaller datasets. However, as discussed in \S \ref{subsec:inefficiency_problem}, when data loading significantly overshadows training time, the benefits of such overlapping become marginal and less effective. Moreover, such data prefetch scheduling usually needs extensive modification on the underlying software infrastructure like the file system (i.e, NoPFS) and data processing system (i.e, ExoShuffle).

\begin{figure}
\centering
\includegraphics[width=\linewidth]{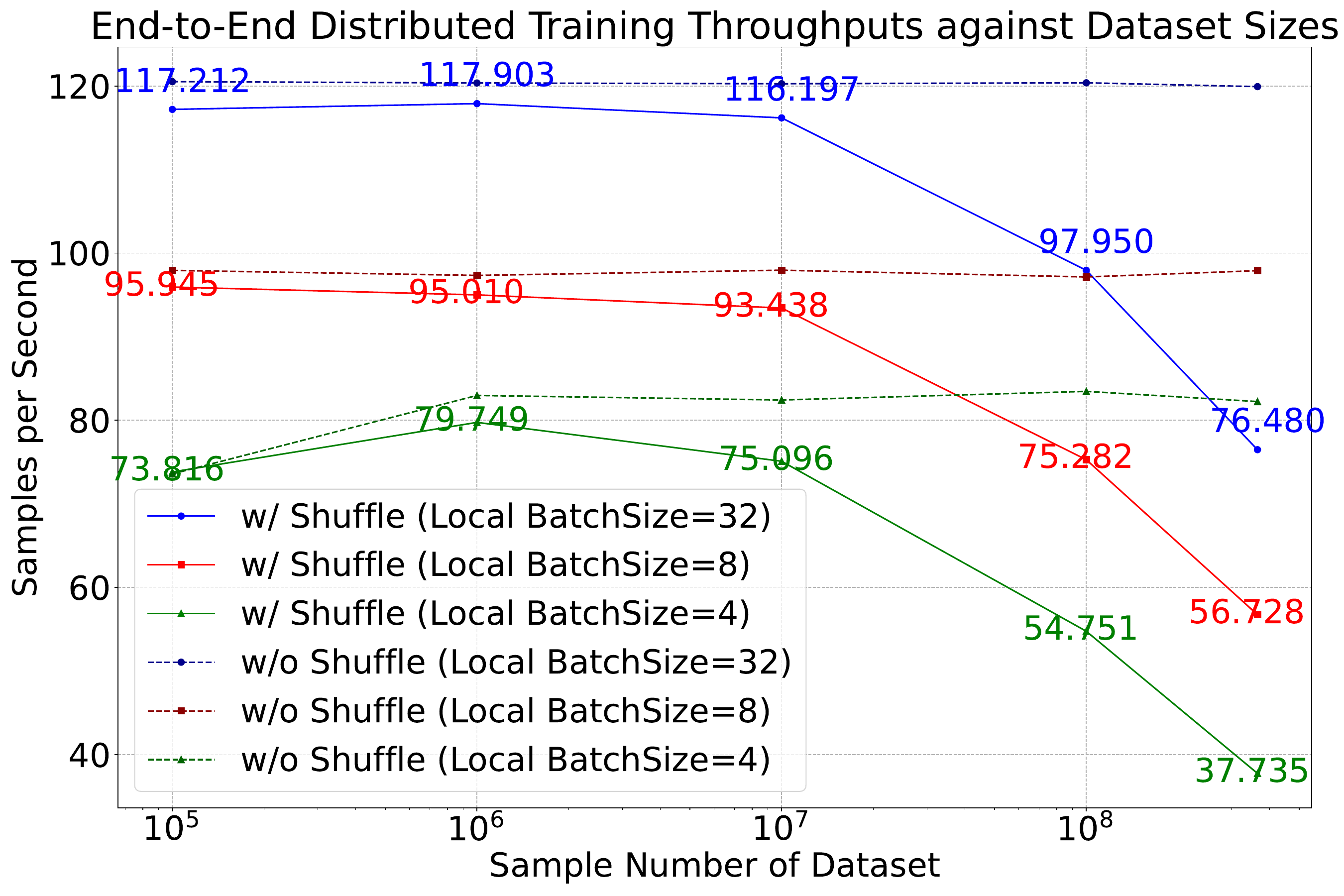}
\caption{Illustration of distributed training throughput for the RoBERTa-base model on various subsets of the C4 dataset, differentiated by batch size. 4 NVIDIA A100 (80\,GB).}
\label{fig:train_throughput_moti_dist}
\end{figure}

\noindent{\bf Data-parallel training.} 
By replicating learners across multiple GPUs, each learner has its own data loading process thus achieving parallel dataset indexing in the global view \cite{parameter_server, li2020pytorch_distributed}. 
To explore the effectiveness of data-parallel training with large shuffled datasets, we conduct the distributed version of the end-to-end training throughput experiment, shown as Figure \ref{fig:train_throughput_moti_dist}. We can see that although data-parallel training can improve the global end-to-end training throughput effectively compared to non-distributed training, data-parallel training still suffers from training throughput degradation when dataset size increases. This is because each learner's data loading procedure is still degraded at larger datasets and hence becomes the bottleneck of its training process, resulting the end-to-end throughput degradation like non-distributed training.
Moreover, utilizing multiple GPUs or hardware resources can be costly, both in terms of initial investment and operational expenses. Also, distributing data and aggregating results across GPUs introduce a communication overhead. This can sometimes offset the training throughput benefits of parallel processing \cite{nguyen2022globally, dp_communication}.

%% file: content/design.tex
\section{ \design{} Design}
\label{sec: design}

In this section, we first show the design principle of \design{} as a new paradigm (\cref{design:paradigm}).
Then, we introduce the design goals of this work (\cref{design:goals}), followed by an analysis of rethinking the randomness in the general learning process (\cref{subsec:insight}). 
Next, we describe the novel execution model under unordered batch generation (\cref{subsec:batchgen_model}) and its requirements on dataset representation and indexing interface (\cref{subsec:ds_interface}).
Finally, this section demonstrates the end-to-end system overview of \design{} (\cref{design:overview}).

\subsection{Towards Overcoming Throughput Degradation: A Paradigm Shift}\label{design:paradigm}

Our approach aims at addressing the critical challenge of loading large shuffled datasets and the resultant throughput degradation inherent in indices mapping. 
By leveraging the deep learning-specific domain knowledge, our approach unleashes unexplored parallelism in the data loading process.

By accelerating the data loading with indices mapping, one of the key benefits is \textit{simplicity}; programmers don't need to bother to solve the above issues in existing solutions described in \cref{motivation:existing-solutions}. Practitioners can confidently apply global shuffling to their model training process to maximize converged accuracy without the concern of training slowdown caused by indices mapping.
This key distinction sets our work apart from existing solutions, focusing on a fundamental shift in how intra-batch data is loaded for batch generation during the training process.

% \jiechen{It is unclear if this sentence: how exactly data is previously handled, and now you shift the data handling in which way else?}.

Moreover, since our method is under the framework of indices mapping, the integration with existing model training environments is straightforward.

In essence, \design{} proposes a paradigm shift in how data is prepared and managed at batch generation stage to achieve dataset shuffling. 
\design{} paves the way for a more easy-to-manage and efficient deep learning process with high accuracy. 
Next, we discuss \design{}'s design goals in detail.

\subsection{Design Goals}\label{design:goals}

Our design objectives are twofold:

\begin{itemize}
\item \textbf{Performance at scale.} Mitigate the data loading bottleneck of indices mapping: the extensive disk random I/Os due to frequent, non-contiguous data indexing,  which can significantly hamper performance. This is an even more challenging goal for TB-scale datasets since the random dataset indexing overhead is positively correlated with dataset size (recall \cref{subsec:inefficiency_problem}).
\item \textbf{Agnostic to the learning process.} \design{} has to reserve and guarantee the global randomness of shuffling for model training process, which can make sure that our approach is not specific to datasets or models.
\end{itemize}

The first goal addresses the performance bottleneck challenge we aim to overcome, while the second goal serves as a constraint to ensure that our solution remains versatile, and capable of being seamlessly incorporated into existing deep learning systems for a variety of tasks. 
To achieve these goals, \design{} leverages the deep learning-domain knowledge obtained by rethinking the randomness in the general learning process. 

\subsection{Rethinking Intra-Batch Sample Randomness}
\label{subsec:insight}

A conventional principle in data loading is to adhere strictly to the order of shuffled datasets. This practice is commonplace in data processing and analysis, ensuring complete randomness in the data presented to the model. In model training, the concept of batch size is introduced to specify the number of samples processed in a single iteration of model training.

The update of the model parameters in one training step can be expressed as:

\begin{equation}
\theta_{\text{new}} = \theta_{\text{old}} - \eta \cdot \nabla_\theta \mathcal{L} \left( \frac{1}{N} \sum_{i=1}^{N} \ell (x_i, y_i; \theta_{\text{old}}) \right)
\end{equation}

In this equation, $\theta$ represents the model parameters, $\eta$ is the learning rate, $\nabla_\theta \mathcal{L}$ denotes the gradient of the loss function $\mathcal{L}$ with respect to the parameters, and $\ell$ is the per-sample loss function. $x_i$ and $y_i$ are the input and target of the $i$-th sample in the batch, and $N$ is the batch size.

We can see that the loss is computed as the average of individual sample losses, suggesting that the order of samples within the same batch does not influence the outcome. This leads us to a deep learning-specific insight: the intra-batch sample order does not impact the learning outcome, opening potential avenues for optimization in data loading and training efficiency. 

To be specific, this insight releases the design of the data loading pipeline from strictly respecting sample order randomness to an intra-batch sample unordered manner. 
This enables additional parallelism space of intra-batch data indexing. 
By out-of-order retrieval, we can parallelize the intra-batch data indexing without worrying about the order of data arrival.

\subsection{Unordered Batch Generation}
\label{subsec:batchgen_model}
Based on the previous observation on model training procedure, we can conceptualize \textit{unordered batch generation}, a versatile data loading pipeline that permits batch generation with unordered intra-batch sample retrieval.

\begin{figure}[t]
\centering
\includegraphics[width=\linewidth]{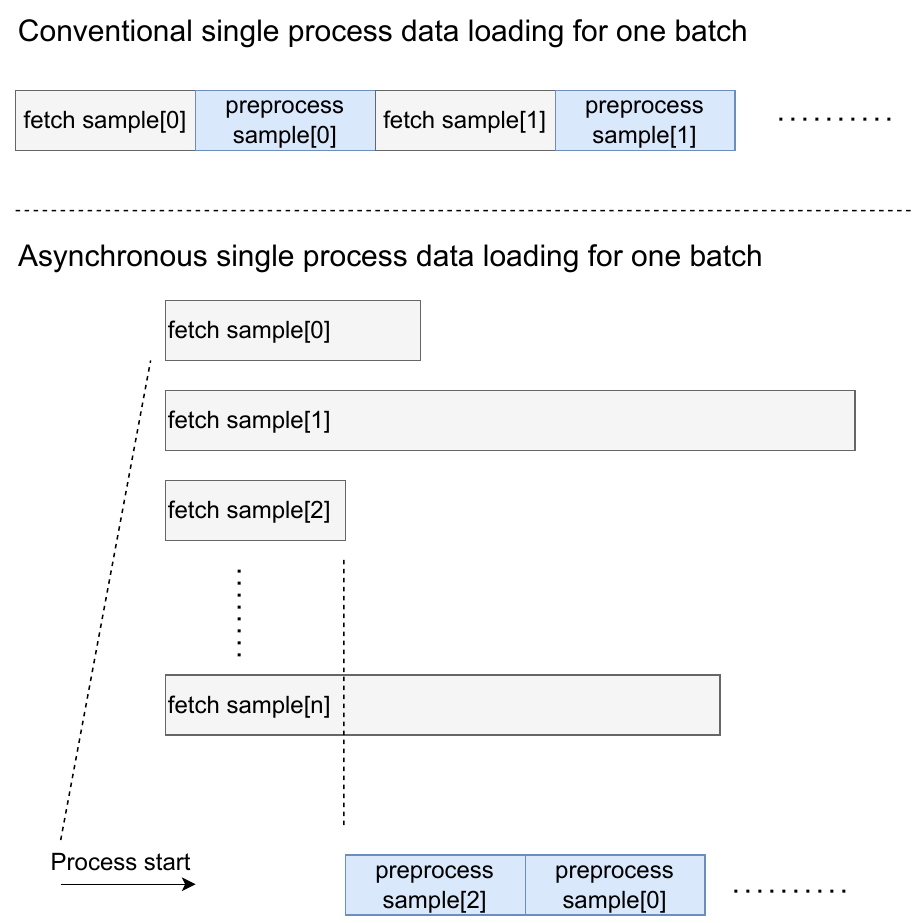}
\caption{A comparative illustration of execution models between conventional and \design{}'s unordered batch generation.}
\label{fig:async_fetch}
\end{figure}

Figure \ref{fig:async_fetch} presents the execution model of both conventional method and unordered batch generation. There are two major parts of unordered batch generation: \textit{parallel dataset indexing} and \textit{overlapped preprocessing}. 

\noindent{\bf Parallel dataset indexing.} By eliminating the necessity to maintain intra-batch sample order, data retrieval operations for distinct samples within a batch can be executed in parallel through asynchronous threading. The data arrival order is changed with such execution but would not affect the learning outcome (recall \cref{subsec:insight}).

\noindent{\bf Overlapped preprocessing.} After data is fetched from the disk, it usually needs to go through the user-defined preprocessing pipeline before being passed into the model training process. Taking preprocessing as a part of the batch generation process, we can safely overlap data preprocessing tasks with data retrieval, thereby achieving additional performance enhancements.

\subsection{Dataset Representation and Indexing Interface}
\label{subsec:ds_interface}

The proposed execution model based on unordered batch generation comes with three requirements on dataset representation and indexing interface.

\noindent{\bf Indexable dataset representation.} Since \design{} is under the framework of indices mapping, the dataset representation should be indexable, which distinguishes \design{} from the non-indexable iterative-style datasets representation like PyTorch iterative datasets and Ray Data \cite{moritz2018ray}. This enables the on-demand indexing to retrieve an arbitrary sample of the dataset at any time.

\noindent{\bf Interference-free retrieval.} The unordered batch generation necessitates the parallel execution of dataset indexing, which requires the data retrieval process to be interference-free with each other. 
If the dataset indexing procedure prevents another dataset indexing procedure from running concurrently, it would fully force the retrieval process back to the one-by-one manner, eliminating the benefits of unordered batch generation.

\noindent{\bf Reusing existing facility.} Considering that there are already many dataset abstractions satisfactory for the above requirements like map-style image datasets in TorchVision, \design{} can directly employ them seamlessly. 
For unsatisfactory datasets, we provide a case study in \cref{sec:impl} to demonstrate how to convert them into the needed fashion.

\subsection{End-to-end View of \design{}}\label{design:overview}

\begin{figure}[t]
\centering
\includegraphics[width=0.6\linewidth]{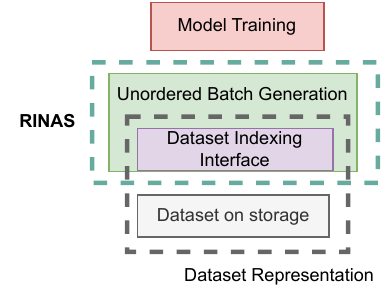}
\caption{A system overview of \design{}. }
\label{fig:sys_overview}
\end{figure}

We now describe the end-to-end view of our approach. Figure \ref{fig:sys_overview} provides a system overview of \design{}. The dataset representation against the dataset storage provides a dataset indexing interface for data retrieval which serves as a \textit{data plane}. \design{}'s unordered batch generation serves as a \textit{control plane} to parallelize the given data retrieval for intra-batch data.

\noindent{\bf Dataset initialization stage.} Data representation needs to contain the mapping from indices to actual data locations. Such information can be stored externally and read at the initialization stage or created at runtime, depending on the storage method of the datasets. The dataset representation also exposes a dataset indexing interface as the data plane to be leveraged by the control plane which is the unordered batch generation module.

\noindent{\bf Batch generation stage.}  The unordered batch generation module executes parallel dataset indexing for data within the same batch and pipelines the preprocessing and IO of different samples. Finally, a batch is generated and fed into the model training process.

\noindent{\bf Scope of \design{}.} Unlike other existing solutions \cite{ray, exoshuffle}, \design{} is not designed for general dataset shuffling applications beyond model training due to the fact that \design{} relies on the deep learning-specific insight (recall \cref{subsec:insight}). Any application that requires the intra-batch sample order to be aligned with the shuffled indices is out of \design{}'s scope.

%% file: content/implementation.tex
\section{Implementation}
\label{sec:impl}

We implement a prototype of \design{} by extending the PyTorch framework and HuggingFace Datasets library with $\sim$400 lines of Python code. 
Our prototype involves an unordered batch generation control plane and an on-demand, parallelizable data plane, each working as a standalone module.

Specifically,  we override the \texttt{\_MapDatasetFetcher} class to enforce the unordered batch generation (\cref{subsec:batchgen_model}), and an asynchronous thread pool is created to fetch data samples in parallel. Note that the index order is changed according to the intra-batch fetch scheduling. Once the data sample is fetched, it is immediately sent to the user-defined preprocessing pipeline, and different data samples are processed in parallel.

\subsection{A Case Study: Converting HuggingFace Datasets}

We provide a case study for how to convert the dataset representation unsatisfactory to our control plane in the needed fashion. The example case here is the HuggingFace datasets, which are based on Apache Arrow format \cite{pyarrow}, a columnar memory format for flat and hierarchical data, organized for efficient analytic operations on modern hardware like CPUs and GPUs.

\noindent{\it Dataset storage.} Specifically, HuggingFace stores datasets as memory-mapped arrow stream files for efficient stream processing \cite{arrow_flight}. Arrow stream format partitions the whole dataset into small data chunks on the storage and enables users to efficiently iterate through the dataset at the unit of data chunks. However, the arrow stream file format lacks data chunk indices which is necessary for index-based loading. To address this, HuggingFace chooses to iterate through the entire dataset at dataset initialization to create a table that contains all the metadata and locations of data chunks on the storage. Unfortunately, this method leads to two major drawbacks:

\begin{enumerate}
    \item[(1)] Long dataset initialization time: the iterating procedure at dataset initialization has a time cost that is linear to the dataset size. For a dataset that is around 1 TB, it can take up to 10 minutes to finish the dataset initialization. 
    \item[(2)] Frequent page swaps during shuffled loading. The dataset initialization process creates a memory-mapped arrow table instance in DRAM, hiding data chunk access details from dataset developers. The data chunk accessing and mapping is fully managed by the operating systems. Due to the limited DRAM space compared to the dataset size, most data chunks are paged out during dataset initialization and need to be re-mapped into DRAM at runtime, which causes frequent page swaps and blocks the parallel access to data chunks.
\end{enumerate}

\noindent{\it Format conversion.} Therefore, we need to optimize the HuggingFace implementation to support the on-demand and parallelized data indexing. Specifically, we convert the files from the arrow stream format to an arrow indexable format, utilizing the \texttt{PyArrow} library \cite{pyarrow}.

\begin{figure}[t]
\centering
\includegraphics[width=\linewidth]{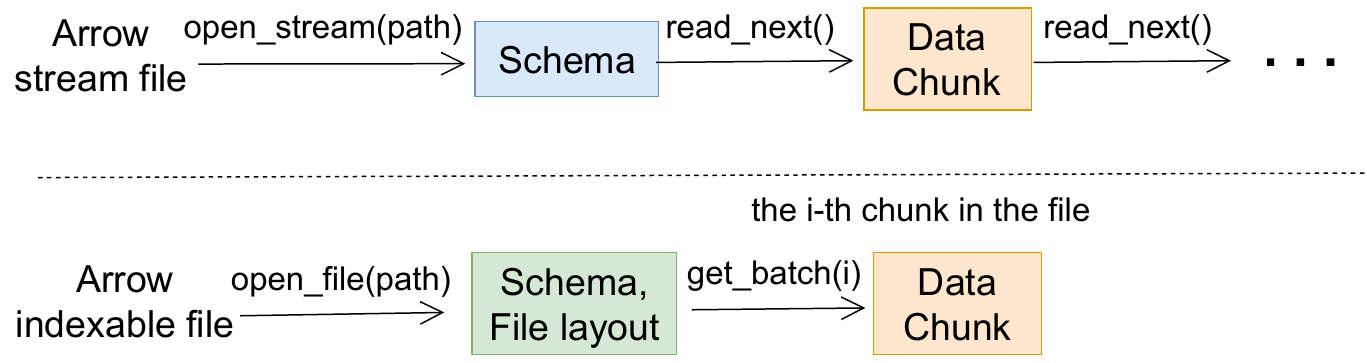}
\caption{A comparative illustration of arrow stream and indexable formats.}
\label{fig:arrow_format}
\end{figure}

Figure~\ref{fig:arrow_format} presents the difference between accessing data chunks in arrow stream file format and arrow indexable file format. The arrow stream file format opens a message stream and sends data chunks in a sequence of messages. The first message is the data schema, which contains the metadata for data chunks in the file. The stream reader iterates through data chunks by \texttt{read\_next()} method without knowing the locations of data chunks. 
On the other hand,  when opening the file, the arrow indexable file format in our prototype reads the data schema containing metadata, and file layout containing data chunk locations. With file layout loaded at the beginning of file reading, the indexable format allows accessing data chunks by its index in the file with \texttt{get\_batch(index)}.

With this format, the data chunks can be accessed without page swaps or memory re-mapping since the data chunk access and mapping becomes on-demand and manageable, facilitating parallel reading of data chunks and aligning with the efficient and scalable batch generation execution model in \design{} (recall~\cref{subsec:batchgen_model}).

\noindent{\bf How to implement \design{} over other deep learning frameworks?} 
In addition to PyTorch, \design{} can be implemented on various deep learning frameworks \cite{abadi2016tensorflow, jax2018github, paszke2019pytorch}. 
Specifically, We provide some tips to enforce \design{} on TensorFlow and JAX.

TensorFlow \cite{abadi2016tensorflow} utilizes the \texttt{tf.data} module for the data input pipeline, but it does not support indices mapping. Therefore, to implement \design{}, first, we need to modify the \texttt{Dataset.batch} function to create the batched shuffled indices. After the shuffled indices are available for TensorFlow datasets, we can force the \texttt{tf.data.Iterator} to iterate through the dataset based on the shuffled indices. When the iterator iterates at the level of batches, the intra-batch data can be fetched in parallel to achieve \design{}'s unordered batch generation (\cref{subsec:batchgen_model}).

Since JAX can directly leverage PyTorch DataLoader for the data input pipeline, our implementation can be naturally supported by adapting our modifications on PyTorch DataLoader.

%% file: content/evaluation.tex
\section{Evaluation}
\label{sec: eval}

We now present the evaluation of \design{} by answering the following questions. 

\begin{enumerate}
    \item[(1)] What's the improvements \design{} provides on the end-to-end model training throughput? (\cref{subsec: throughput_eval})
    \item[(2)] How do the control and data plane contribute to \design{}'s performance in various scenarios? (\cref{subsec:analysis})
    \item[(3)] How does the global randomness brought by \design{} benefit the learning outcome? (\cref{subsec:learning_benefits})
    \item[(4)] What kinds of overhead \design{} introduces and how much is it? (\cref{subsec:overhead})
\end{enumerate}

\noindent{\bf Testbed.}
We set up a testbed on a standard computing node equipped with an AMD EPYC 7742 CPU with 96 GB RAM and 4 NVIDIA A100 GPUs. Each GPU has 80\,GB memory.
In addition, we rely on the cluster-wide WEKA file system for dataset storage \cite{weka}, which is commonly adopted in real-world and supported by high-demand computing platforms like Amazon AWS and Google Cloud. For software configurations, we utilize CUDA 11.7, PyTorch 2.0.1, HuggingFace Datasets 2.14.6.dev0, and Transfomers 4.34.0.dev0. 

\noindent{\bf Methodology.}
We evaluate \design{} using two datasets: C4 and ImageNet, which are typical text and image datasets for large language models and computer vision models. Correspondingly, we train a typical large language model RoBERTa and a computer vision model ResNet-152 to demonstrate \design{}'s benefits.
To evaluate the end-to-end training throughput, we conduct a series of model training iterations, spanning 300 steps. The throughput is calculated by dividing the total number of samples processed during these iterations by the total time taken to complete them. The results are averaged over 3 runs. To avoid startup interference, we run enough warm-up rounds before collecting the results.

\noindent{\it Large language model.}
We train a RoBERTa-base model based on the C4 dataset. RoBERTa (Robustly Optimized BERT Approach) is an advanced variant of the BERT model, and the cleaned English branch of C4 dataset has $\sim$305\,GB raw text data and $\sim$1.1\,TB after being tokenized by RoBERTa.
We compare \design{} against HuggingFace, which is the default approach to store datasets and execute language model training \cite{liu2019roberta, devlin2019bert, touvron2023llama}.

\noindent{\it Computer vision model.} We train the ResNet-152 model using the ImageNet dataset that has $\sim$140\,GB raw image data. The baseline is the training pipeline that employs the official PyTorch DataLoader and the ImageNet dataset implementation by TorchVision \cite{torchvision2016}, which is widely used for computer vision models \cite{simonyan2015deep, he2016deep, dosovitskiy2021image}. 

Additionally, we also test the baselines when supercharged by Ray for both text and image model training. Notice that we could not convert the used datasets with Ray Data (ExoShuffle) due to the extensive DRAM usage for such transformation (\cref{subsec:overhead}). The only option is to use Ray Train's wrappers for HuggingFace and PyTorch DataLoaders.

\subsection{End-to-end Training Throughput}
\label{subsec: throughput_eval}

\begin{figure}[t]
\centering
\includegraphics[width=\linewidth]{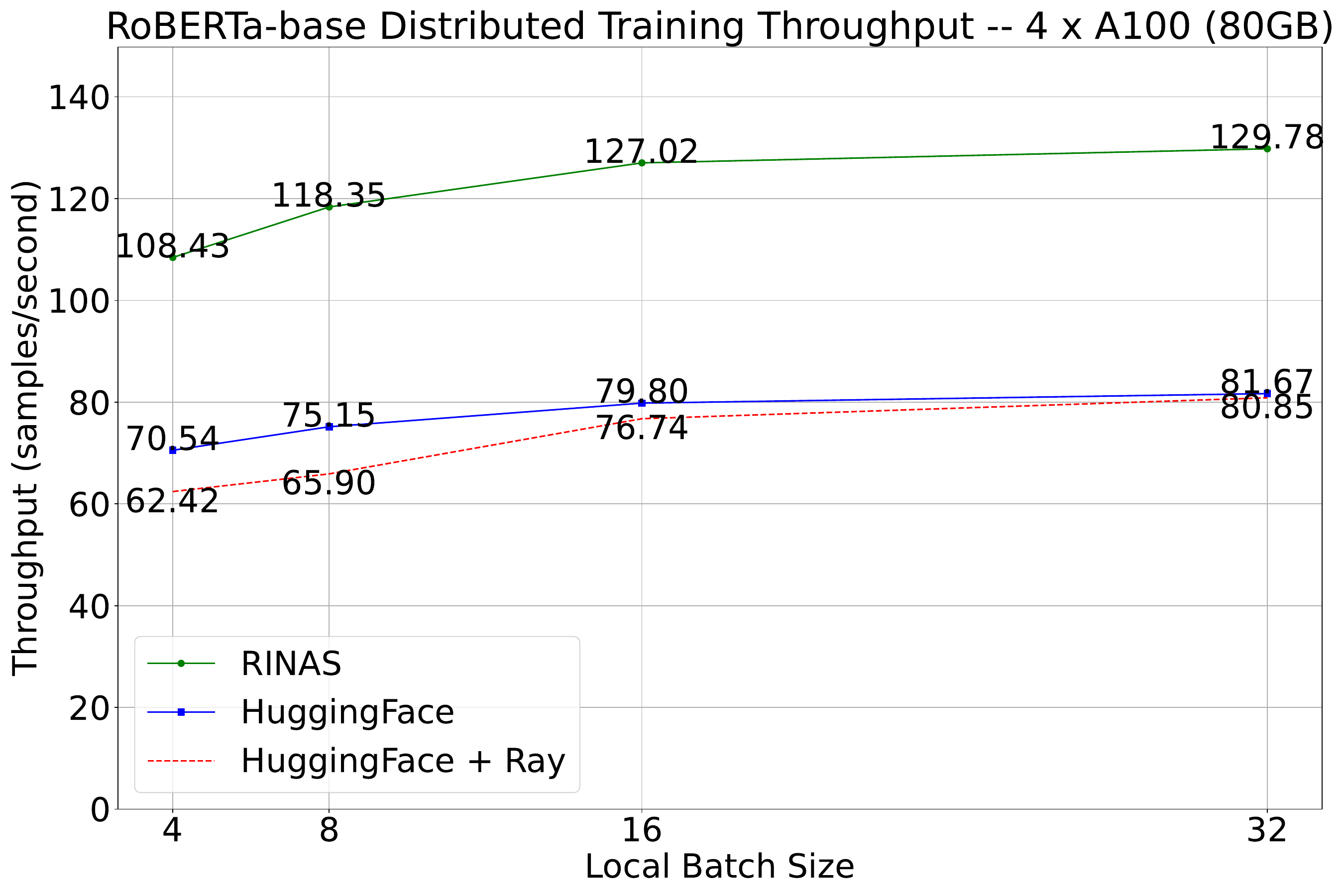}
\caption{ RoBERTa-base distributed training throughput comparison between HuggingFace and \design{}.
}
\label{fig:roberta_train_throughput_dist}
\end{figure}

\noindent{\bf RoBERTa-base training upon C4}. Figure \ref{fig:roberta_train_throughput_dist} presents the end-to-end distributed training throughput of the RoBERTa-base model when the batch size increases. We can see a general trend where \design{} can achieve larger throughput at larger batch sizes; this is due to the higher parallelism of \design{}'s unordered batch generation at larger batch sizes. 

\begin{figure}[t]
\centering
\includegraphics[width=\linewidth]{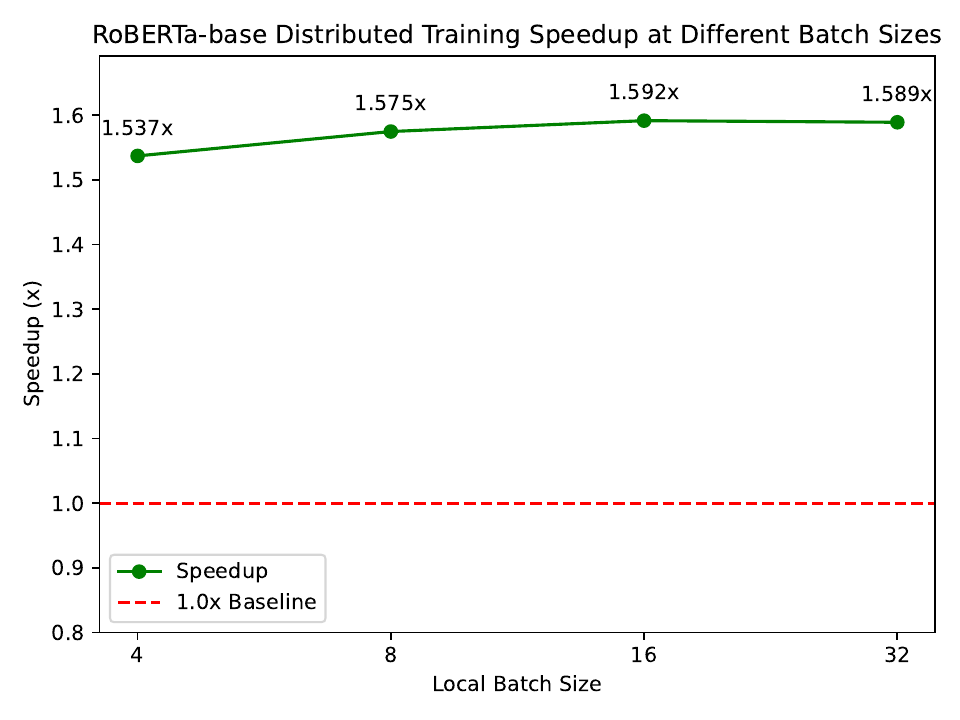}
\caption{ RoBERTa-base distributed training speedup over HuggingFace at different batch sizes.
}
\label{fig:train_speedups_dist}
\end{figure}

To further analyze the performance gain provided by \design{}, we measure the training throughput of employing original HuggingFace pipeline as a baseline and show the speedup of \design{} at different batch sizes as the Figure~\ref{fig:train_speedups_dist} shows.
As Figure~\ref{fig:train_speedups_dist} shows, \design{} can improve the end-to-end training throughput by 1.54-1.59$\times$, demonstrating the efficiency of \design{} at a wide range of batch sizes. 
\design{} achieves higher speedup ratios at larger batch sizes, suggesting \design{}'s better scalability than HuggingFace in terms of batch sizes.

In addition, when the HuggingFace baseline is supercharged by Ray, the training throughput has a slight drop. This is because Ray only takes the data loading process as its actor's application workload and does not optimize the inner pipeline. With additional overhead from Ray's processes, it is reasonable to see a slight performance drop.

\begin{figure}[t]
\centering
\includegraphics[width=\linewidth]{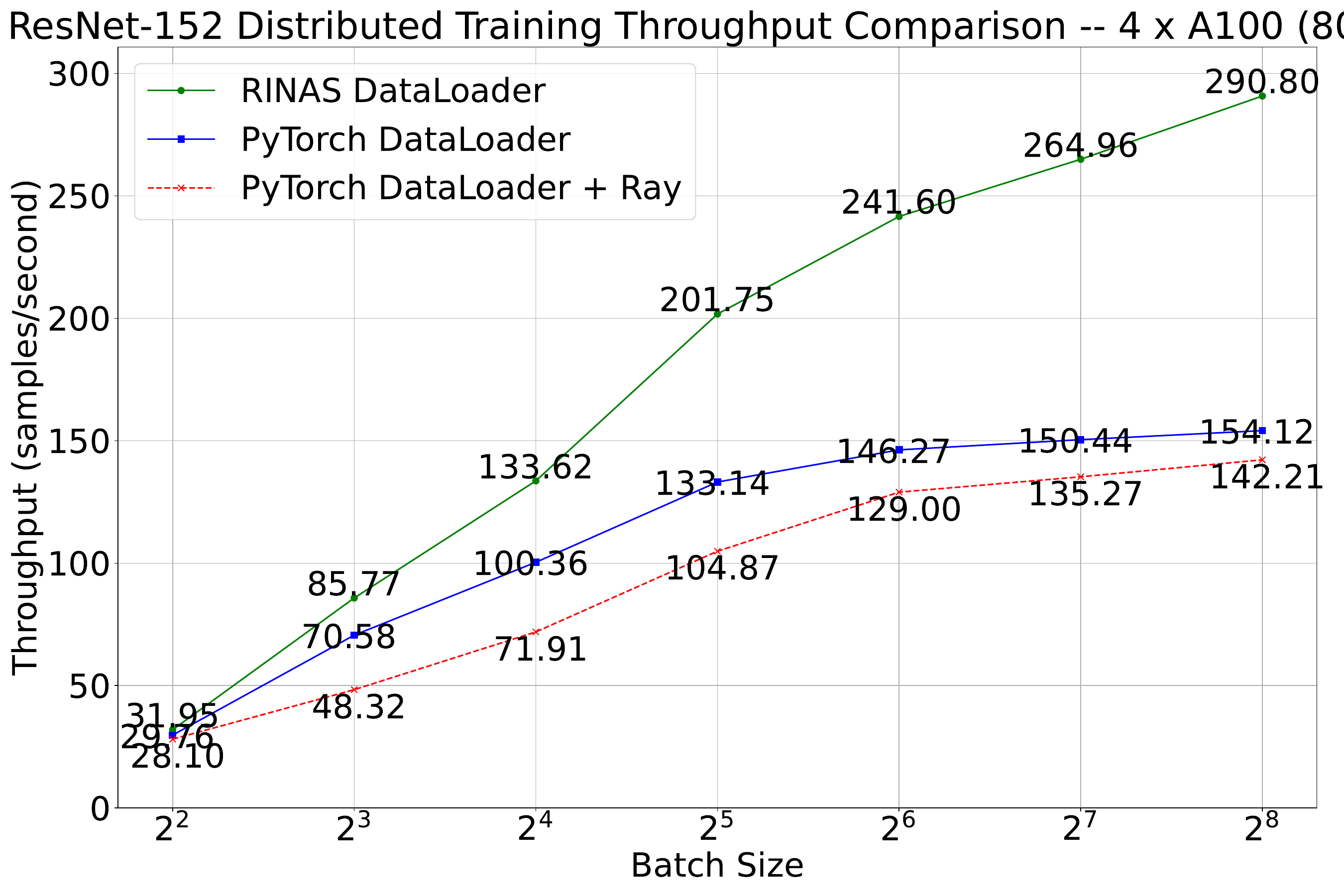}
\caption{ResNet-152 distributed training throughput comparison between using PyTorch dataloader and \design{}.
}
\label{fig:resnet_train_throughput_dist}
\end{figure}

\begin{figure}[t]
\centering
\includegraphics[width=\linewidth]{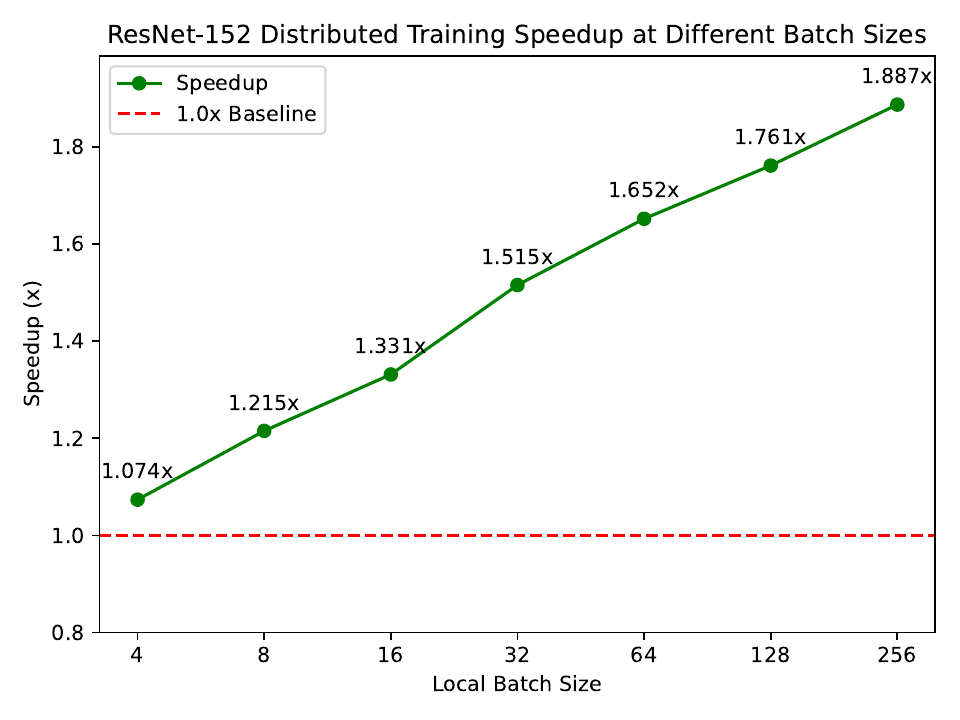}
\caption{ ResNet-152 distributed training speedup at different batch sizes.
}
\label{fig:resnet_speedups_dist}
\end{figure}

\noindent{\bf ResNet-152 training on ImageNet}. Figure \ref{fig:resnet_train_throughput_dist} presents the end-to-end distributed training throughput of the ResNet-152 model when the batch size increases.
We can observe the same trend with RoBERTa training: \design{} can achieve higher training throughput when batch size increases at larger batch sizes. For comparison, training throughput using PyTorch DataLoader remains at a relatively low level, suggesting the effectiveness of \design{} in image datasets like ImageNet.

To further analyze the performance gain by \design{}, Figure~\ref{fig:resnet_speedups_dist} shows the speedup ratios compared to the PyTorch DataLoader baseline at different batch sizes.
Similar to Figure \ref{fig:train_speedups_dist}, We can also see that \design{} can achieve higher speedup ratios by up to 1.89$\times$ at larger batch sizes, demonstrating \design{}'s scalability in the image dataset in terms of batch sizes as well.

We can also notice the slight performance drop when PyTorch DataLoader is supercharged by Ray which is due to the same reason as explained in the language model training experiment.

\subsection{Breakdown Analysis}
\label{subsec:analysis}

\begin{figure}[t]
\centering
\includegraphics[width=\linewidth]{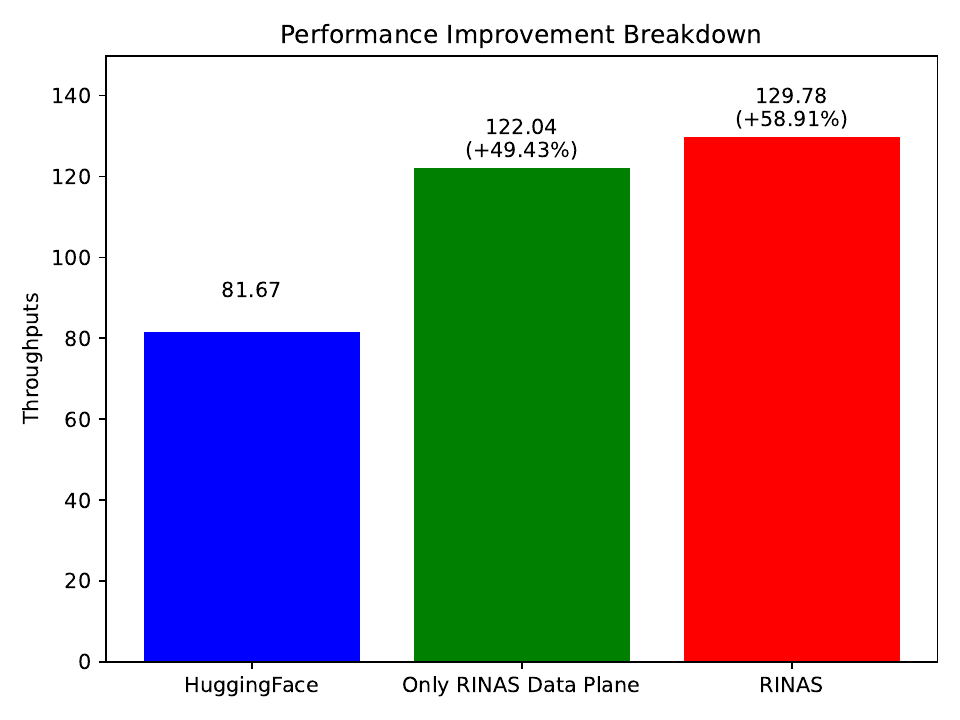}
\caption{RoBERTa training throughput improvement breakdown under a batch size of 32.}
\label{fig:improve_breakdown}
\end{figure}

In order to analyze the benefits of \design{}'s control plane and data plane, we analyze their contributions to the aforementioned experiments: On one hand, we measure the RoBERTa-base end-to-end distributed training throughput at a batch size of 32 by disabling the control plane to showcase the contribution from solely data plane. Note that removing the control plane does not require any modification on the side of the data plane for \design{}. Figure~\ref{fig:improve_breakdown} displays the results. We can see that \design{}'s data plane can improve RoBERTa-base distributed training throughput by 49.4\% when the batch size is 32. When the control plane is further enforced, we can achieve a 58.9\% cumulative performance improvement, highlighting the great contribution from the \design{} data plane. 
On the other hand, in the case of ResNet-152 training on the ImageNet dataset, both \design{} and PyTorch DataLoader operate on the same dataset implementation, differing only in their respective control planes within the DataLoader. This highlights the importance of \design{} control plane in image datasets like ImageNet. As for the speedup ratios difference between text and image datasets, we would like to 

\begin{table}[t]\small
\begin{tabular}{c|c|cc|l}
\hline
\multirow{2}{*}{Model}     & \multirow{2}{*}{Dataset}                  & \multicolumn{2}{c|}{Acc. w/ shuffle}         & \multirow{2}{*}{Imps.} \\
                           &                                           & \multicolumn{1}{c|}{limited}    & global     &                        \\ \hline
TabNet                     & HIGGS (7.5 GB)                            & \multicolumn{1}{c|}{$\sim$70\%} & $\sim$76\% & 1.09$\times$           \\ \hline
\multirow{3}{*}{ResNet-50} & DeepCAM (8.2 TB)                          & \multicolumn{1}{c|}{$\sim$78\%} & $\sim$83\% & 1.06$\times$           \\
                           & ImageNet-21k (1.1 TB)                     & \multicolumn{1}{c|}{$\sim$37\%} & $\sim$45\% & 1.22$\times$           \\
                           & \multicolumn{1}{l|}{ImageNet-1k (140 GB)} & \multicolumn{1}{c|}{$\sim$50\%} & $\sim$70\% & 1.40$\times$           \\ \hline
ResNet-18                  & criteo (1.3 TB)                           & \multicolumn{1}{c|}{$\sim$30\%} & $\sim$90\% & 3.00$\times$           \\ \hline
VGG-19                     & criteo (1.3 TB)                           & \multicolumn{1}{c|}{$\sim$20\%} & $\sim$90\% & 4.50$\times$           \\ \hline
\end{tabular}
\caption{Model training accuracy discrepancies between limited shuffle and global shuffle.}
\label{table: accuracy}
\end{table}

\subsection{Convergence Benefits}
\label{subsec:learning_benefits}

As \design{} guarantees the global shuffling against the datasets, global shuffling benefits all gradient-decent-based trainers based on the theoretical foundation \cite{meng2019convergence}.

Due to the extensive long time to train a model on large datasets until convergence, Table \ref{table: accuracy} collects and presents the model accuracy discrepancies discovered by other studies~\cite{nguyen2022globally, exoshuffle, xu2022stochastic}. The studied models contain TabNet~\cite{arik2020tabnet}, ResNet, and VGG and the studied datasets include HIGGS~\cite{misc_higgs_280}, ImageNet, DeepCAM~\cite{nguyen2023deepcam} and criteo~\cite{criteo}. We can observe that the accuracy discrepancies vary a lot depending on different models and datasets; no matter what models and datasets the benefits of global shuffling are always obvious. Besides vision models, in practice tabular data and models such as language models and temporal series models are more sensitive to the shuffle quality~\cite{tabular_shuffle, ray_global_shuffle, schluter-varab-2018-data}, further highlighting the necessities of global shuffling in model training. 

\subsection{Resource Overhead}
\label{subsec:overhead}

We consider \design{}'s CPU memory usage overhead at dataset preparation and both CPU and GPU memory overhead at runtime.

\noindent{\bf Dataset preparation}. For PyTorch Datasets, \design{} can directly optimize the shuffled loading performance upon PyTorch Datasets. Thus, there is no additional memory usage overhead for PyTorch Datasets. As for \design{}'s file format transformation required for HuggingFace datasets, our transformation process is based on the PyArrow stream processing which only requires less than $\sim$100 MB DRAM capacity.

For comparison, Ray Data module provides a conversion method from PyTorch and HuggingFace Datasets to Ray Datasets, which can be stored and shuffled in Ray's shared-memory object store. However, their implementations do not allow us to finish such a transformation for either ImageNet or C4 on our testbed, due to the out-of-memory errors caused by extensive memory usage of such a transformation. 
This results from the fact that Ray needs to load the entire dataset into its local object store before shuffling \cite{exoshuffle}, while \design{} follows an indices mapping manner and only loads data into RAM on demand.

\noindent{\bf Runtime}. At runtime, another CPU memory usage overhead comes from the additional thread pool employed by \design{}'s control plane implementation, which relies on the multi-threaded asynchronous intra-batch data fetching. 

Since each intra-batch data is mapped and fetched by a separate thread, the number of threads needs to match the batch size, which requires a large number of threads to create when the batch size increases. In the case of distributed ResNet-152 training with a batch size of 256 on 4 GPUs described in Sec.~\cref{subsec: throughput_eval}, the entire system needs to create a thread pool of 256 threads on each learner's process, cumulatively 1024 threads at a time in total. 

This can potentially stress the CPU and the storage system while other data loaders without such design typically only require a single thread for each data loader process. However, considering that the modern training clusters are typically equipped with many-core CPUs and usually not fully exploited during model training, \design{}'s additional stress on the CPU can be handled.

Also, since the parallelism degree of \design{}'s unordered batch generation is positively correlated with batch size, a larger batch size per GPU is necessary to have a significant speedup which leads to larger CPU and GPU memory consumption. However, based on the theoretical foundation \cite{gao2020study} and the results of practice \cite{KANDEL2020312} , a large batch size benefits model's generalization ability.  Due to this reason, it has been a common practice to pursue larger batch sizes for model training with various memory optimization techniques \cite{korthikanti2022reducing, zhao2023fsdp, chen2016training}.

%% file: content/discussion.tex
\section{Discussion}
\label{sec: discussion}
This section discusses some immediate concerns one may have about \design{}.

\noindent{\bf Does \design{} provide high-level dataset representations?}
\design{} introduces a new tradeoff between the performance benefits and the dataset representations in its implementation. Ideally, a unified dataset representation can facilitate the programming for datasets, such as the memory-mapped Arrow table in HuggingFace, with which developers don't need to know where the dataset is located. However, to gain \design{}'s performance benefits (recall ~\cref{sec: eval}), the dataset developers should manipulate the data retrieval process, majorly involving locating the data segments on the storage (recall ~\cref{subsec:ds_interface} and~\cref{sec:impl} ).

\noindent{\bf Is \design{} able to be extended to other distributed computing systems?}
\design{} introduces a novel shuffled loading paradigm for batch generation models. The key operation is the unordered batch generation, which mainly relies on asynchronous random IO to parallelize the dataset indexing (recall~\cref{subsec:batchgen_model}). Therefore, it is feasible to extend \design{} by enforcing unordered batch generation by integrating the asynchronous random IO into the dataset iterators supported by a variety of data processing systems, such as Ray, Spark, and Twister \cite{ray, 10.1145/1851476.1851593, spark}. However, since \design{} is designed with deep learning-specific domain knowledge (recall~\cref{subsec:batchgen_model}), the use of \design{} needs the consideration of whether their applications are also applicable.

\noindent{\bf How should \design{}'s performance to be expected on even larger datasets?} In general, the size of the dataset to be used for training should be aligned with the model size. To investigate \design{}'s real benefits with larger dataset sizes, we also need to adjust the model we train. It would be very interesting to see how it performs on larger datasets to train a truly large foundation model, but it is beyond our capacity in computational resource. Investigating such problem can be one of our future directions.

\section{Related Work}

\noindent{\bf Shuffling in data processing systems. } Since MapReduce \cite{dean2008mapreduce} and Hadoop \cite{hadoop}, there are plenty of solutions with a focus on optimizing disk I/O efficiency and pipelining for in-place shuffle operation in data processing systems \cite{iShuffle, hadoop_shuffle, ownership}.  However,  under model training workloads, prior approaches are usually computationally heavy, and hard to be pipelined (recall ~\cref{sec: background}). 

To address this issue, Ray~\cite{ray} and ExoShuffle~\cite{exoshuffle} propose a distributed futures-based shuffle system to provide better flexibility and interoperability with model training workloads. However, those two proposals mandate loading the entire dataset into its object store before they read and shuffle the data, causing infeasible demands on DRAM resources for large datasets (recall~\cref{subsec:overhead}).

\noindent{\bf Shuffling in deep learning systems. } Due to the inefficiency of shuffled loading with indices mapping in large datasets (recall ~\cref{subsec:inefficiency_problem}), the programmers may have to use buffered shuffling at the cost of model convergence accuracy. This inspires some deep learning systems to modify the global access order and the buffer eviction scheme accordingly~\cite{sun2022solar, DeepIO_buffer_shuffle} to maximize the data reuse and the buffer hit rate. However, similar to partial shuffling, this breaks the true randomness of global shuffling no matter modifying inter-batch or inter-epoch order, which limits the applicable scope to specific models and datasets. Similar to NoPFS~\cite{10.1145/3458817.3476181}, they usually also involve significant efforts in modifying the underlying software infrastructure.

Notably, all these solutions for reducing shuffle-related overheads in model training can leverage \design{}'s unordered batch generation to further enhance the performance. 
Our insight is obtained from an observation of the general model training process.

\noindent{\bf Shuffle pattern exploration for ML. } Some researchers work on exploring theoretical convergence bound of different local shuffle patterns \cite{meng2019convergence, Nguyen2020AUC, Ahn2020SGDWS, Rajput2021PermutationBasedSI}, with a hope to find an alternative to the expensive global shuffling. However, their analysis is usually limited to convex problems with low-dimensional data, leaving their effectiveness on real-world problems and datasets not fully explored.

\noindent{\bf Data loading parallelism. }
Previous proposals also leverage multi-core parallelism to uncover data loading parallelism in different types of systems. 
First, while our paper focuses on faster data loading from local SSDs, some works focus on faster data loading from a remote datastore or file system through the network. 
Barclay et al.~\cite{barclay1994loading} leverage dataflow parallelism for memory-to-memory data loading via parallel TCP connections.   
Yang et al.~\cite{yang2019accelerating} propose to accelerate data loading for PyTorch-based distributed training systems from network-attached remote storage. 
Second, some other works focus on local data loading, the same as this work, leveraging asynchronously I/O to unleash unordered parallelism on in-memory~\cite{lim2014mica,zhao2022altocumulus} or on-disk databases~\cite{cheng2014scanraw,dziedzic2017dbms}. 
ScanRaw speculatively uses more cores to load data when additional disk bandwidth is available~\cite{cheng2014scanraw}. 
Dziedzic et al.~\cite{dziedzic2017dbms} have a similar observation of this paper that file format is important on data loading performance and draws the conclusion that data loading is CPU-bound. 
We inherit previous work's insights on leveraging asynchronous I/Os to solve an unexplored problem: releasing the data loading order of each batch in model training.

\noindent{\bf Preprocessing acceleration. } Some researchers are focusing on the preprocessing acceleration for higher batch generation efficiency. For example, NVIDIA Data Loading Library (DALI) \cite{DALI} accelerates data loading for image datasets by moving the image preprocessing workload to GPU. Since we focus on different stages of the data loading pipeline and thus orthogonal to each other,  our work can be combined with DALI when working with image datasets.

\section{Conclusions}

Loading shuffled large datasets has become the key bottleneck in model training throughput. However, existing solutions face different limitations and usually leave programmers to face the trade-off between training speed and convergence accuracy. 
In this work, we propose \design{}, a data loading framework with the design of a novel execution model with unordered batch generation. By leveraging the deep learning-specific domain knowledge, \design{} unleashes the additional parallelism space in the shuffled loading process.
Our evaluation shows that \design{} accelerates model training substantially under the guarantee of global randomness in shuffling.
This work does not raise any ethical issues.